\documentclass{emulateapj}
\usepackage{graphicx}
\def\lsim{\lower.5ex\hbox{$\; \buildrel < \over \sim \;$}}
\def\gsim{\lower.5ex\hbox{$\; \buildrel > \over \sim \;$}}

\def\t{\ifmmode {\tau} \else $\tau$ \fi}

\def\ref{\noindent \hangafter=1 \hangindent=0.7 truecm}

\def\cm{\ifmmode {\rm cm}^{-1} \else cm$^{-1}$ \fi}
\def\s{\ifmmode {\rm s}^{-1} \else s$^{-1}$ ./clusters/cl1226\fi}
\def\cc{\ifmmode {\rm cm}^{-3} \else cm$^{-3}$ \fi}
\def\cs{\ifmmode {\rm cm}^{-2} \else cm$^{-2}$ \fi}
\def\g{\ifmmode \gamma \else $\gamma$\fi}
\def\G{\ifmmode \Gamma \else $\Gamma$\fi}

\def\kms{\ifmmode {\rm km\ s}^{-1} \else km s$^{-1}$\fi}

\begin{document}
 
\title{Month-Timescale Optical Variability in the M87 Jet}

\author{Eric S. Perlman\altaffilmark{1,2}, D. E. Harris\altaffilmark{3},
John A. Biretta\altaffilmark{4}, William B. Sparks\altaffilmark{4}, 
F. Duccio Macchetto\altaffilmark{4,5}}

\altaffiltext{1}{Joint Center for Astrophysics, Physics Department, 
University of Maryland, Baltimore County, 1000 Hilltop Circle, Baltimore, MD,
21250, USA.  Email:  perlman@jca.umbc.edu}
 
\altaffiltext{2}{Department of Physics and Astronomy, Johns Hopkins University,
3400 North Charles Street, Baltimore, MD, 21218, USA}                         

\altaffiltext{3}{Smithsonian Astrophysical Observatory, 60 Garden Street,
Cambridge, MA,  02138, USA}

\altaffiltext{4}{Space Telescope Science Institute, 3700 San Martin Drive,
Baltimore, MD,  21218, USA}   

\altaffiltext{5}{Space Telescope Division of the European Space Agency, ESTEC,
Noordwijk, Netherlands}       


\begin{abstract}

A previously inconspicuous knot in the M87 jet has undergone a dramatic
outburst and now exceeds the nucleus in optical and X-ray luminosity.
Monitoring of M87 with the {\it Hubble Space Telescope} and {\it Chandra} X-ray
Observatory during 2002-2003, has found month-timescale optical variability in
both the nucleus and HST-1, a knot in the jet $0.82''$ from the nucleus. We
discuss the behavior of the variability timescales as well as spectral energy
distribution of both components.  In the nucleus, we see nearly
energy-independent variability behavior.  Knot HST-1, however, displays weak
energy dependence in both X-ray and optical bands, but with nearly comparable
rise/decay timescales at 220 nm and 0.5 keV. The flaring region of HST-1
appears stationary over eight months of monitoring.  We consider various
emission models to explain  the variability of both components.  The flares we
see are similar to those seen in blazars, albeit on longer timescales, and so
could, if viewed at smaller angles, explain the extreme variability properties
of those objects.

\end{abstract}

\section{Introduction}

M87 is among the nearest galaxies with a bright radio/optical/X-ray jet.  
Its proximity (distance 16 Mpc, Tonry 1991, giving
a scale of 78 pc/arcsec) allows features to be studied with unparalleled
spatial resolution.

The dynamic nature of the M87 jet was first  recognized in the X-rays by Harris
et al. (1997, 1998), who found variability on  $\sim 1$ year timescales. 
In the optical, Biretta et al. (1999) found superluminal motion throughout the
inner $10''$ of the jet, with speeds up to  $6c$, as well as gradual changes in
flux. These observations did not explore timescales $\lsim 6$ months, however. 
More recently, the jet's optical spectrum was found to harden in knots, 
consistent with local particle acceleration (Perlman et al. 2001a).  And {\it
Chandra} observations found spectral indices   $\alpha_x > 1$  $(S_{\nu}
\propto \nu^{-\alpha})$, broadly consistent with an extrapolation of the
radio-optical synchrotron emission (Marshall et al. 2002, Wilson \& Yang
2002).  X-ray synchrotron emission implies radiative lifetimes $\sim$ 1-10
years, and thus requires {\it in situ} acceleration.  

All this evidence suggested the possibility of shorter-timescale variability 
in the M87 jet.  Indeed, recently Harris et al. (2003, hereafter paper I)
detected X-ray flaring on timescales $\sim 1 $ month during 2002 in both the
nucleus and HST-1, a knot in the jet. 
Here we discuss observations of the M87 jet with both {\it HST} and
the {\it Chandra} X-ray Observatory, which find flares in both 
the nucleus and knot
HST-1 on timescales of $\sim 2$ months.
A second paper (Biretta et al. 2003, hereafter paper III)
discusses the longer timescale variability of the M87 jet.

\section {Observations and Data Reduction}

We make use of {\it Chandra} and {\it HST} observations of M87 during November
2002 - June 2003.  During this interval we observed M87  roughly every 20-40
days.  Table 1 lists the  {\it HST} observations; the {\it Chandra} data will
be discussed in an upcoming paper (Harris et al., in prep.).  

Our {\it Chandra} data reduction procedures were detailed in Paper I. All {\it
HST} observations were reduced in IRAF and PyRAF with the best available flat
fields, biases, darks and illumination correction images.  Dithered images were
combined using PYDRIZZLE, which combines drizzling (Fruchter \& Hook 2002) with
cosmic ray and hot pixel removal, geometric correction and mosaicing. Galaxy
subtraction was done on the F475W and F814W data using ELLIPSE, BMODEL and
IMCALC.  Identical galaxy models were used at all  epochs.    Flux-calibrated
images were obtained using SYNPHOT.  We applied a standard Galactic extinction
law and assumed $N(H)=2.4 \times 10^{20} {\rm cm^{-2}}$ to correct the fluxes
for reddening.  Polarized light images were combined into Stokes' parameters in
AIPS, using standard formulae.

Considerable effort was expended to estimate uncertainties.
We accounted for several error sources, including
zero-point and slope errors in the SYNPHOT calibration, flat-fielding
errors, and Poisson errors in the data and in modelling and
subtraction of the galaxy.  Errors were propagated by adding in
quadrature.  The cumulative error in the {\it HST} photometry is
typically 3-4\%, and is dominated by the flat-fielding and
zero-point errors.

Lightcurves were extracted from our data at three UV/optical
wavelengths - 220 nm, 475 nm and 814 nm, and three X-ray bands (paper
I) - 0.2-0.75 keV (``soft''; nominal 0.5 keV), 0.75-2 keV (``medium'';
nominal 1.4 keV), and 2-6 keV (``hard''; nominal 4 keV). The 2003
March 31 and 2003 May 10 (part of a snapshot program led by D. Maoz)
images were taken at a slightly longer wavelength; to place them on a
common 220 nm flux scale we assumed $\alpha_o=0.6$ (Perlman et
al. 2001a).  Finally, we have also made use of flux points from
2001-2002 {\it HST} data which are discussed in more detail in Paper
III.

\section {Results}

Figure 1 shows the 2003 April 17 F220W image of the inner $4.5''$ of the M87
jet,  as well as optical and X-ray lightcurves for the nucleus and HST-1 
respectively.  As can be seen, both components vary on
timescales of a few months in the optical, with typical month-to-month
variations being about 10-15\% at 220 nm.   
Other regions do not vary significantly from month to month and so are
not shown in Figure 1.  For the nucleus, flux was extracted from $0.2''$ and
$0.6''$ square boxes, which showed identical variability.  In Figure 1 we use
the  $0.2''$ box.  For knot HST-1, we extracted flux from a $0.6''$ box.  The
extraction regions for our {\it Chandra} data were described in paper I. Our
data confirm Tsvetanov et al.'s  (1998) claim of month-timescale nuclear optical
variability.  However, this is the first such report for any jet component.   

\begin{deluxetable}{ccc}
\tablewidth{0pt}
\tablecolumns{3}
\tablecaption{Log of HST Observations}
\tablehead{
 \colhead{Date} & \colhead{Instrument \& Bands} & \colhead{Program}}
\startdata
2002-11-30 	& ACS F220W, F475W, F814W & 9705\\
2002-12-07 	& ACS F606W + POLVIS & 9705\\
2002-12-22	&ACS  F220W, F475W, F814W & 9705\\
2003-02-02	& ACS  F220W, F475W, F814W & 9705\\
2003-03-06	&  ACS F220W, F475W, F814W & 9705\\
2003-03-31	& ACS F250W, F330W & 9454 \\
2003-04-17	& ACS F220W, F475W, F814W & 9705\\
2003-05-10	&ACS  F250W, F330W & 9454 \\
2003-06-07	& STIS F25QTZ & 9474 
\enddata
\end{deluxetable}

%
%
%

\begin{figure}
\centerline{\includegraphics*[scale=0.45]{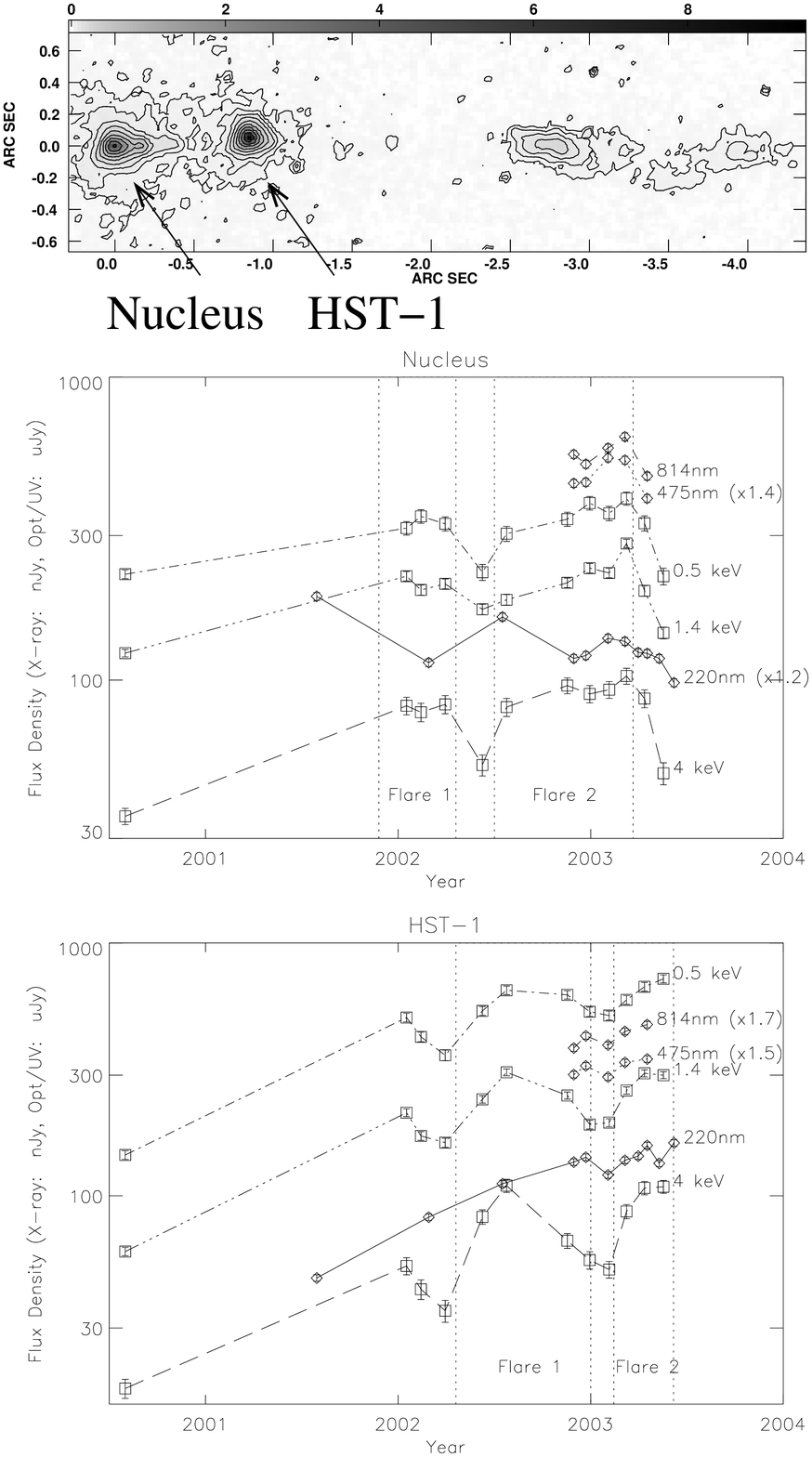}}  

\caption[]{At top, we show the 2003 April 17 HST/ACS F220W image of the 
inner $4.5''$ of the M87 jet. The image has been rotated so that the
jet appears along the {\it x}-axis.  The middle and bottom panels show
lightcurves for the nucleus and HST-1.  Some of the optical/UV
lightcurves are multiplied by arbitrary factors to separate them
from the X-ray lightcurves. Flare events discussed in the text are
enumerated.  All error bars are at the $1\sigma$ level.}

\end{figure}

At 220 nm, $>95\%$ of the flux in both varying regions is contained within a 
distribution consistent with the PRF.  Thus both are  unresolved, and so
$<0.02''$ (1.5 pc) in size.  This implies light-travel times within a factor
3-20 (depending on the Doppler factor $\delta=[\Gamma(1-\beta \cos
\theta)]^{-1}$) of the variability timescales (\S4, Paper I).  The nucleus's
variable region is $<0.01''$ from the galactic center,  with HST-1's varying
region being $0.82''$ (64 pc) distant, locations identical to $\pm~ 0.01''$ of
those measured with {\it Chandra} (Paper I).  We detect no motion in HST-1's
flaring region;  therefore, the flare likely occurred in the nearly stationary
component at its upstream end (Biretta et al. 1999), rather than one that
propagates at the velocities seen in moving components within HST-1, namely
$v_{app} = 6c$ ($0.013''$/year, easily detectable  in these data).


Figure 2 shows the F606W polarized light image. As can be seen, the nucleus is
essentially unpolarized.  HST-1 shows fractional
polarization $P=0.46$ at its flux peak, a maximum $P=0.68$ (nearly the
theoretical maximum for synchrotron radiation from an ordered magnetic field)
at its  upstream end $\sim 0.72''$ from the nucleus, and a minimum $P=0.23$
$0.92''$ from the nucleus.  The magnetic field vectors in HST-1 are
perpendicular to the jet direction, consistent with a shock.  The alternative
interpretation of a tightly wound helix would require cycles separated by
$<0.02''$ to be consistent with the observed morphology. A much higher bulk
$\Gamma$ would also be needed (because of the longer path) to be consistent
with $v_{app}=6c$ (Biretta et al. 1999).  The polarization at HST-1's flux
peak is much higher than in 1995 ($P=0.14$, Perlman et al. 1999). This is in
line with the properties of BL Lac objects, which often have higher optical
polarizations in high states (e.g., Hagen-Thorn et al. 1998).

\begin{figure}

\centerline{\includegraphics*[angle=270,scale=0.35]{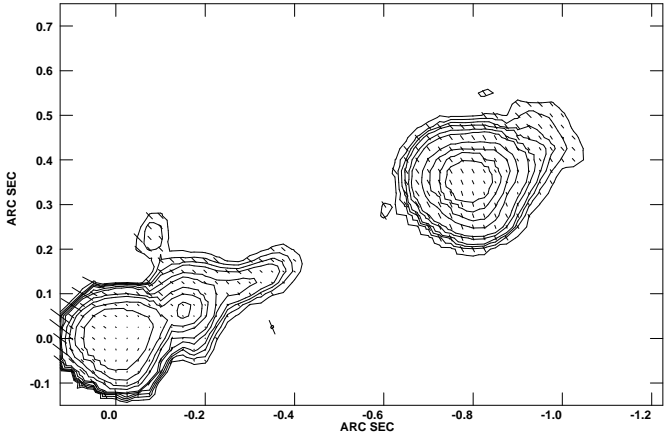}}

\caption[]{An image of the innermost $1.5''$ of the M87 jet, with polarization
vectors (magnetic field) superposed.  The magnitude of the vector is
proportional to  the percent polarization, with a $0.02''$ vector representing
100\% polarization.
Contours are shown at ( 0.2, 0.4, 0.6, 0.8, 1, 1.5, 2, 4, 6, 10, 20) ADU/sec}

\end{figure}


\section {Constraints from the Lightcurves}

To constrain the physics, we compute doubling and halving timescales $\tau_2,
\tau_{1/2}$, given in Table 2 for timeperiods referenced in Figure 1.  We can
compare these with predictions from models with $E^2$ losses, which are of the
form $\tau\propto \nu^\eta$.   If the emission is synchrotron radiation and
synchrotron cooling dominates, $\eta=-0.5$, (Kirk, Mastichiadis \& Rieger 1998;
Paper I).  As shown in B\"ottcher et al. (2003),  if synchrotron self-Compton
(SSC) losses dominate, the dominant factor in $d\gamma/dt$ is the synchrotron
radiation energy density $U_{sy}$, so that $\eta= (q-4)/2$ = --1 to --0.75 for 
electron spectral index $q=2.5$ to 2 [$N(\gamma) \propto \gamma^{-q}$, which is
related to $\alpha$ via $\alpha=(q-1)/2$]. 

\begin{deluxetable*}{llllcccccc}
\tablewidth{0pt}
\tablecolumns{10}
\tablecaption{Flare Timescales}
\tablehead{
& & & & \multicolumn{6}{c}{Doubling or Halving Timescale in Band (days)} \\
\colhead{Component} & \colhead{Band} & 
\colhead{Dates} & \colhead{Type (Flare \#)} & \colhead{814 nm} &
\colhead{475 nm} & \colhead{220 nm} & \colhead{0.5 keV} & \colhead{1.4 keV} &
\colhead{4 keV}
}
\startdata
Nucleus & X-rays & June-July 2002 & Rise (2) & ... & ... & ... & $135 \pm 39$ &  $608 \pm 474$ & $83 \pm 20$  \\
Nucleus & X-rays & March-May 2002 & Fall (1) & ... & ... & ...& $78 \pm 11$ & $71 \pm 6$ & $67 \pm 9$   \\
Nucleus & Optical & Dec. 2002-Feb. 2003 & Rise (2)& $317 \pm 108$ & $206 \pm 47 $ & $ 256 \pm 82$ & ... & ... & ... \\ 
Nucleus & Optical & Mar.-Apr. 2003 & Fall (2) & $81 \pm 12$ & $83 \pm 12$ & $274 \pm 106$&  ... & ... & ...  \\
HST-1 & X-rays & Feb.-Apr. 2003 & Rise (2) & ... & ... & ... & $223 \pm 49$ & $118 \pm 13$ & $ 61 \pm 9$  \\
HST-1 & X-rays & Nov.-Dec. 2002 & Fall (1) & ... & ... & ... & $149 \pm 55$ & $91 \pm 17$ & $54 \pm 10$ \\
HST-1 & Optical & Feb.-Apr. 2003 & Rise (2) & $359 \pm 82$ & $ 414 \pm 108$ & $243 \pm 36$ & ... & ... & ... \\
HST-1 & Optical & Dec. 2002 - Feb. 2003 & Fall (1) & $260 \pm 133$ & $217 \pm 91$ & $139 \pm 37$& ... & ... & ... \\
\enddata
\end{deluxetable*}

Table 2 shows several differences between the variability behavior of the
nucleus and HST-1.   The nucleus does not show strong energy-dependence
in either the X-rays ($\eta_{x,rise}=-0.24 \pm 0.21$ and $\eta_{x,fall}=-0.07
\pm 0.10$) or optical ($\eta_{o,rise}=-0.09 \pm 0.43$ and $\eta_{o,fall}=0.52
\pm 0.32$, but the departure from zero in the latter measurement is due
entirely to the lower significance of the fall in F220W during 2003 Feb.-Apr.;
further, note that the drop accelerated after monitoring in F475W and
F814W ceased).   By comparison, HST-1
does show energy-dependent behavior in the X-rays, with $\eta_{x,rise}=-0.62
\pm 0.13$ and $\eta_{x,fall}=-0.49 \pm 0.21$. But in the optical, our data
are less constraining, giving  $\eta_{o,rise}=-0.35 \pm 0.22$ and
$\eta_{o,fall}=-0.51 \pm 0.78$.

HST-1's behavior in the X-rays agrees with the predictions of a simple
synchrotron model.  But such a model cannot easily account for the comparable
optical and X-ray variability timescales.  Two other models can explain such
behavior.  The first is that the flare was caused by adiabatic  compression,
followed by expansion on a dynamical timescale of  $\tau_{dyn}\sim 200$ days. 
This would allow radiative losses dominate at high energies, where  $\tau_{syn}
<  \tau_{dyn}$, while expansion losses would dominate at low energies. The
X-ray variability behavior then requires $B \approx 2 ~\delta^{-1}$ mG. 
However, this model has difficulty explaining X-ray and optical increases and
decreases that do not exactly coincide in time (Figure 1).  Alternately, the
X-ray flare could be triggered by shock compression (as suggested by the
polarimetry, \S 3), with the optical emission representing the shocked plasma's
downstream propagation.  This model requires a $\sim 10\times$ stronger
magnetic field downstream of the shock to explain the similar 0.5 keV and 220
nm variability timescales.

The nearly energy-independent variability behavior of the nucleus  disagrees
with simple synchrotron models, although some cannot be
excluded  formally because of the large error bars. 
One could account for the energy independence either by adiabatic
compression and expansion, or a helical trajectory, where flux would change
with viewing angle  (suggested by Urry et al. 1993 to explain the nearly
energy-independent variability of PKS 2155$-$304 in 1991).   But perhaps the
most attractive model is that the varying region has $R/c \approx 70 ~\delta$
light-days, so that the light-crossing time controls the variability behavior
(e.g., Chiaberge \& Ghisellini 1999).  In this case we would require 
$B\lsim 20 ~\delta^{-1}$ mG.

\section {Comparison to Blazar Flares}

The flares of the nucleus and HST-1 (Figure 2) resemble those of blazars,
albeit on longer timescales.
Can we explain blazar flares ($\tau \sim 0.2-5$ days; Ulrich, Maraschi \& Urry
1997; Pian 2002), as highly beamed versions of these events?  Timescales
Lorentz-transform as $\tau^\prime \propto \tau \delta^{-1}$, so $\tau=5$ days
requires $\delta_{M87}/\delta_{blazar}\approx 1/12$ (also consistent
with the luminosities seen in blazar flares, e.g., Giebels et al.
2002), while $\tau=1$ day
requires $\delta_{M87}/\delta_{blazar}\approx 1/60$.  
For reasonable $\Gamma$, 
this requires $\delta_{M87} \lsim 2$, at the lower end of the
range considered for HST-1 in Paper I,   and implies $\delta_{blazar} =
20-100$.   Consistency with HST-1's $v_{app}=6c$ (Biretta et
al. 1999), 
then requires $\Gamma > 10$ and $\theta = 15-19^\circ$,  much
tighter constraints than in previous works.  
A timescale  
$\tau=5$ days then 
requires $\Gamma=10$ and $\theta=1^\circ$ if $\delta_{M87}=1.6 ~
(\Rightarrow \Gamma_{M87}=10, \theta_{M87}=19^\circ$), while $\tau=1$ day
requires $\Gamma>25$ and $\theta<1^\circ$, plus smaller $\delta_{M87}$. 
Unified schemes usually constrain $\Gamma$ and $\theta$ less tightly (e.g., 
$\Gamma= 5-30$ and $\theta < 10-30^\circ$, Urry \& Padovani 1995).  However,
one can imagine a range of $\delta$ or stronger magnetic fields (e.g., Li \&
Kusunose 2000, Kataoka et al. 2000, B\"ottcher et al. 2003).

Blazars can show either energy-dependent or energy-independent
behavior, even within the same campaign.  Often, 
energy-dependent behavior includes one band leading the other.
With our data, 
any discussion of delays is premature, although there are
possible indications in the data.  Monitoring of M87 continues, and we
will discuss this subject in a
later paper.


\begin{figure}

\centerline{\includegraphics*[scale=0.45]{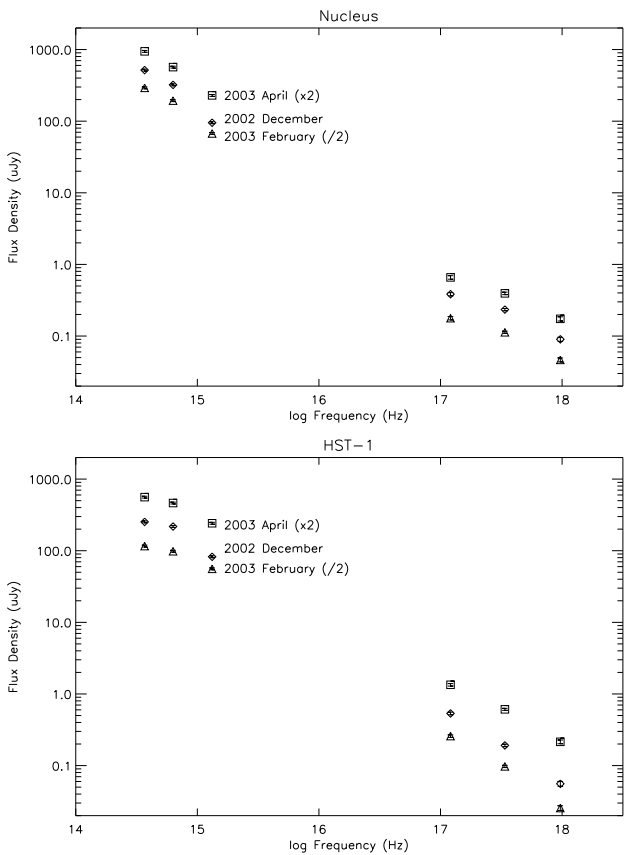}}

\caption[]{Spectral Energy Distributions for the nucleus (top) and
knot HST-1 (bottom).  Three epochs are shown:  2002 December, 2003
February, and 2003 April.}

\end{figure}

\section{Spectral Variability}

Do we also see spectral variability in these components?   Figure 3 shows
broadband spectra in three epochs, representing both high and low states.

The nucleus has $\alpha_o>1$ and $\alpha_{ox}>1$, in agreement with
the fit value of $\nu_{break} \approx 3 \times 10^{12}$ Hz found using
historical data (Perlman et al. 2001b).  But in all epochs, $\alpha_x<
\alpha_{ox}<\alpha_o$, where $\alpha_o, \alpha_{ox}$ and $\alpha_x$ are, 
respectively, the spectral indices in the optical, optical-X-ray and
X-ray bands.  Thus for standard synchrotron models, the X-ray emission
must come from a component distinct from that responsible for the optical
emission, since a single electron population cannot produce spectral
hardenings blueward of $\nu_{break}$.  The nucleus displays only
modest spectral variability, with the 2002 December points (a low
optical state) having flatter $\alpha_o$ but steeper $\alpha_{ox}$.
This is consistent with a significant SSC contribution.

By contrast, HST-1 has  $\alpha_o < \alpha_{ox} < \alpha_x$ in all epochs, with
$\alpha_o=(0.41,0.47,0.59)$ for epochs (2003 April, 2002 December, 2003
February), similar to that seen in 1998, when it was $10\times$ fainter in
optical (Perlman et al. 2001a).  These values of $\alpha_o$ are consistent with
shock injection.  The optical to X-ray spectrum  steepens in low states
and flattens in high states, as expected for synchrotron flaring. 
Interestingly, the 2003 April high state has a clearly harder spectrum
than that seen in  2002 December.


\section{Final Thoughts}

Here and in paper I we have found strong optical and X-ray variability in  the
nucleus of M87, and knot HST-1 in its jet.   Is it possible to link the
variability of the nucleus and HST-1 in any way?  Perhaps the flaring in HST-1
was triggered by a density enhancement travelling down the beam of the jet.  If
we assume a near-constant apparent speed of $v_{app}=6c$  for this material,
its ejection would have occurred $\approx 30-35$ years ago.  Such an episode
could plausibly have caused a radio flare in the nucleus. Interestingly,
DeYoung (1971) detected $\sim 30$\% variations in the nuclear radio flux in
1969-1971.  It is possible (but speculative) to link the flaring in the nucleus
in $\sim 1970$ to the  current behavior of HST-1.

\begin{acknowledgments}

We thank the STScI staff for accommodating our director's discretionary time
request on short notice.  We thank O. Stohlman for assistance with {\it
Chandra} data reduction, and C. A. Padgett for assistance with {\it HST} data
reduction. We acknowledge C. Dermer, M. B\"ottcher and A. S. Wilson for
interesting discussions, and D. Maoz for additional {\it HST} data.  We thank
an anonymous referee for comments that improved this work significantly.  E. S.
P. thanks the Smithsonian Astrophysical Observatory for hospitality during a
visit in February 2003.  

E. S. P. acknowledges support from NASA LTSA
grant NAG5-9997 and {\it HST} grant GO-9705.01. Research on M87 at STScI is
supported  by {\it HST} grants GO-7274, GO-8048, GO-8140, GO-8780, GO-9461 and
GO-9474.  Work at SAO was supported by NASA contract NAS8-39073 and grants
GO2-3144X and G03-4124A.  

\end{acknowledgments}

\end{document}